\documentstyle[aps,prl,multicol,amssymb,amsfonts]{revtex}
\title{On the Nonlocality of the Quantum Channel in the Standard Teleportation
Protocol}
\begin{document}

\author{Rob Clifton}
\address{Departments of Philosophy and History and Philosophy of
Science, \\ 10th floor, Cathedral of
Learning, University of
Pittsburgh, \\ Pittsburgh, PA\ 15260, USA. \\ email: rclifton+@pitt.edu}
\author{Damian Pope}
\address{Department of Physics, University of Queensland \\
Brisbane 4072, Queensland,
Australia \\ email: pope@physics.uq.edu.au}
\date{\today}
\maketitle

\begin{abstract} By exhibiting a violation of a novel form of the Bell-CHSH inequality, 
\.{Z}ukowski 
has recently established that the quantum correlations exploited in the standard
perfect teleportation protocol cannot be recovered by any local hidden
variables model. Allowing the quantum channel state in the protocol to 
be given by any density operator of two spin-1/2 particles, we 
show that a violation of a generalized form of
\.{Z}ukowski's teleportation
 inequality can only occur if the channel state, considered by itself, violates a 
 Bell-CHSH inequality.   On the other hand, although it is sufficient 
 for a teleportation process to have a nonclassical fidelity---defined 
 as a fidelity exceeding $2/3$---that the channel state employed violate a Bell-CHSH 
 inequality, we show that such a violation does \emph{not} imply a violation of  
 \.{Z}ukowski's teleportation
 inequality or any of its generalizations.  The
 implication does hold, however, if the fidelity of the teleportation exceeds 
 $2/3(1+1/2\sqrt{2})\approx .90$, suggesting the 
 existence of a regime of nonclassical values of the fidelity, less 
 than $.90$, for 
 which the standard teleportation protocol can be modelled by local 
 hidden variables.
 \end{abstract} \draft

\begin{multicols}{2}

\section{Introduction}
It is well-known that the quantum correlations predicted by pure entangled 
states violate Bell-CHSH inequalities \cite{popescu_rohrlich} and, therefore,  
cannot be recovered by any local hidden variables model. Yet 
these same correlations
cannot by themselves be used to transmit information between the locations
of the entangled systems.  This has sometimes been taken to 
suggest that quantum 
correlations are not inherently `nonlocal', but simply `nonclassical'. 
On the other hand, it is a striking fact that quantum
correlations \emph{can} used, as in quantum teleportation, to increase
the information carrying capacity of a classical channel.  This new 
operational manifestation of quantum entanglement invites a 
deeper analysis of the extent to which the quantum teleportation 
process itself can be modelled classically, and what role, if 
any, nonlocality must play in explaining the success of teleportation.    

 The standard quantum teleportation protocol runs as follows \cite{6man}.
Consider three qubit systems 1, 2, and 3, fix an
arbitrary direction in space to define the $z$-direction for each qubit, 
and let 
$\left| \uparrow \right\rangle $ and $\left| \downarrow
\right\rangle $ denote the eigenstates of $\sigma_{z}$.  
We can imagine that all three qubits are initially in the possession of
someone named Clare, who follows the instructions of Alice and Bob 
to prepare qubits $2+3$
in a quantum channel state given by some density operator $D$ 
of their choice, and prepares qubit $%
1 $ in a pure state $\left| \phi \right\rangle $,
unknown to Alice or Bob. Clare then feeds qubits $2$ and $3$ to Alice and
Bob, respectively, to use as their ``quantum channel'', and Clare passes
qubit $1$ to Alice so that she can teleport its unknown state to Bob. To
execute the protocol, Alice measures the ``Bell operator'' on qubits $1+2$, 
which has eigenstates \cite{bmr}: 
\begin{eqnarray}
\left| \Phi _{\pm }\right\rangle &=&1/\sqrt{2}(\left| \uparrow
\right\rangle \left| \uparrow \right\rangle \pm \left| \downarrow
\right\rangle \left| \downarrow \right\rangle ), \\
\left| \Psi _{\pm }\right\rangle &=&1/\sqrt{2}(\left| \uparrow
\right\rangle \left| \downarrow \right\rangle \pm \left| \downarrow
\right\rangle \left| \uparrow \right\rangle ),
\end{eqnarray}
 and she obtains one of four results $n=1,...,4$ corresponding to the eigenstates
above, with probabilities $p_{n}$. For each outcome $n$, Alice then sends
classical information, via (say) a normal telephone call, instructing Bob to perform a
corresponding unitary transformation $U_{n}$ on his qubit $3$. Having
followed Alice's instructions, Bob's qubit will be left in one of four
states $D_{n}$. 

If the quantum channel state $D$ is a maximally
entangled pure state, the unitaries $U_{n}$ can always be chosen so that no matter 
what Alice's Bell operator measurement outcome, Bob will have prepared his particle by the completion of
the protocol in exactly the same state $D_{n}=\left| \phi \right\rangle
\left\langle \phi \right| $ ($n=1,...,4$) as the state of qubit $1$ that Alice 
teleported.  For example, when $D$ is the singlet state, qubit 3 has 
an equal probability of being in one of the four states 
$\left| \phi \right\rangle\left\langle \phi \right|$, 
$\sigma_{x}\left| \phi \right\rangle\left\langle \phi 
\right|\sigma_{x}$, $\sigma_{y}\left| \phi \right\rangle\left\langle \phi 
\right|\sigma_{y}$, $\sigma_{z}\left| \phi \right\rangle\left\langle \phi 
\right|\sigma_{z}$, so that Bob's performing, respectively, one of the four unitary 
transformations
\begin{eqnarray} \nonumber 
U_{1}=I\ \  (\left| \Psi _{-}\right\rangle), & \ U_{2}=
\sigma_{x}\ \  (\left| \Phi _{-}\right\rangle),\\ \label{standard}U_{3}=\sigma_{y}\ \  
(\left| \Phi _{+}\right\rangle),\ & U_{4}=\sigma_{z}\ \  (\left| \Psi _{+}\right\rangle),
\end{eqnarray} 
suffices for perfect teleportation.

For a mixed and/or non-maximally entangled quantum channel state $D$,
perfect teleportation of an unknown state cannot be achieved (cf., 
e.g., \cite{werner}). However, one
can introduce a natural measure of success for any fixed ``strategy'' 
$n\mapsto U_{n}$ of associating unitary operators $U_{n}$ with Alice's 
Bell operator outcomes.  This measure is 
the fidelity $\mbox{$\mathcal{F}$}_{\{U_{n}\}}(D)$ of
transmission to Bob, given by the uniform average, over all possible unknown
states $\left| \phi \right\rangle $, of the quantity $\sum_{n=1}^{4}p_{n}%
{Tr}(D_{n}\left| \phi \right\rangle \left\langle \phi
\right| )$ \cite{popescu},\cite{gisin}. Taking the supremum of 
$\mbox{$\mathcal{F}$}_{\{U_{n}\}}(D)$ over all possible strategies for associating unitary operators $%
U_{n}$ with Alice's Bell operator outcomes, one obtains a number, 
$\mbox{$\mathcal{F}$}_{max}(D)$, representing the 
\textit{maximum achievable} fidelity of teleportation for a given fixed
quantum channel state $D$ \cite{horodeckis}. A density matrix $D$ is 
then regarded
as useful for nonclassical teleportation just in case $\mbox{$\mathcal{F}$}_{max}%
(D)>2/3$, where $2/3$ is the maximum achievable fidelity without the
quantum channel, or when a classical channel is substituted in its
place \cite{popescu},\cite{massar}.

Let us return, for the moment, to the case of perfect teleportation, where $%
D$ is a maximally entangled pure state and $\mbox{$\mathcal{F}$}_{max}(D)=1$. The
striking thing is that Alice only directly communicates to Bob two
classical bits of information when she conveys to him one of the four integers $%
n=1,...,4$. (We assume, as usual, that they have agreed in advance 
on a strategy $n\mapsto U_{n}$.)  Indeed, one might think that
 Bob's sure-fire reconstruction of the unknown state $\left|
\phi \right\rangle $ as a state of his qubit $3$ 
entails that he
must (somehow) have actually received a \emph{full qubit's worth} of information
(i.e., the information contained in specifying the two of infinitely many
possible real numbers needed to fix the normalized expansion 
coefficients of the state $\left|
\phi \right\rangle $). A natural, though perhaps naive,
explanation would be that the correlations inherent in a maximally entangled
pure state carry the extra information to Bob instantaneously and
nonlocally when that state ``collapses'' as a result of Alice's Bell 
operator measurement. Indeed, the view that the quantum channel itself carries some
share of the net information received by Bob appears to have been favoured
by the inventors of teleportation (\cite{6man}, p. 1896; cf. also \cite
{popescu}, p. 797), and is the springboard for a number of different
explanations of teleportation from the point of view of particular
interpretations of quantum theory (\cite{vaidman}--\cite{bohm}).
There are sceptics, however, who have challenged the view that perfect
teleportation requires nonlocal information transfer of any sort (\cite
{hardy}--\cite{ari}).  Moreover, one should bear in mind that the 
information about the unknown state $\left|
\phi \right\rangle $ ``carried'' by Bob's qubit at the completion 
of a \emph{single} run of the protocol is not actually \emph{accessible} 
to him, 
since it is well-known that there is no way for him to discern the quantum-mechanical state 
of a single system. 

In the case of explaining imperfect nonclassical teleportation, 
where $1>\mbox{$\mathcal{F}$}_{max}(D)>2/3$, the situation is more complicated, and appears to
favour the sceptics. While $D$ must still be entangled (i.e., not a
convex combination of product states of $2+3$), $D$ can still satisfy 
\textit{all} Bell-CHSH inequalities \cite{popescu} 
\begin{equation}
  |\left\langle\bar{\sigma}_{1}\otimes \bigl( \sigma_{1}
+ \sigma _{2} \bigr)
+ \bar{\sigma} _{2}\otimes \bigl( \sigma_{1}
- \sigma _{2} \bigr)\right\rangle_{D}|\leq 2,
\end{equation}
where the $\bar{\sigma}$'s and $\sigma$'s denote arbitrary spin observables
of qubits $2$ and $3$, respectively, and $\left\langle \cdot \right\rangle
_{D}$ denotes expectation value in the state $D$. More precisely,
if we let $\beta (D)\in [2,2\sqrt{2}]$ be the maximum over all
Bell-CHSH expressions of the above form, then $\mbox{$\mathcal{F}$}_{max}(D)>2/3$ does 
\textit{not} imply $\beta (D)>2$, even though the converse implication does
 hold \cite{horodeckis}. So nonclassical teleportation is possible even in
the absence of an independent argument for the nonlocality of the quantum
channel state via Bell's theorem. Notwithstanding this, one could adopt the
view that the condition $\mbox{$\mathcal{F}$}_{max}(D)>2/3$ \emph{itself} should 
be viewed as sufficient
for thinking of the channel state $D$ as nonlocal (cf. \cite{popescu},
p. 799). However, the sceptic is likely to point out that this gives further
support to his view that nonclassical teleportation can occur without the
involvement of nonlocality (cf. \cite{hardy}, p. 6).

All parties to this discussion are agreed that good teleportation is made
possible by the fact that the correlations of an entangled quantum channel
state $D$ are nonclassical. They only diverge on the question of
whether or not the standard teleportation protocol necessarily involves
nonlocality. Recently,
\.{Z}ukowski \cite{z-guy} has made an interesting attempt to force 
this issue
of nonlocality by asking an extremely relevant 
question: Can the correlations of $D$ that are \textit{actually involved%
} in the standard teleportation protocol be recovered in a local hidden
variables model? \.{Z}ukowski attacks this question by making a novel use of
a Bell-CHSH inequality, which we explain in the next section.
The two sections following, Sections \ref{section_three} and 
\ref{section_four}, investigate the
connection between a generalized form of \.{Z}ukowski's teleportation
 inequality and the standard Bell-CHSH inequality
for the channel state. In particular, we show that a violation of the former implies that
the channel state satisfies $\beta(D) > 2$, but that  
$\beta(D) > 2$ does \emph{not} imply a violation of even our generalized form
of \.{Z}ukowski's teleportation inequality. However, in Section \ref{section_five} 
we show that such a violation occurs whenever the fidelity 
$\mbox{$\mathcal{F}$}_{st}(D)>2/3(1+1/2\sqrt{2})\approx .90$, where $\mbox{$\mathcal{F}$}_{st}$ is the 
fidelity associated with the \emph{standard} choice for Bob's unitary 
operators given in (\ref{standard}).   The implications of these results for the 
interpretation of quantum teleportation as nonlocal are discussed in 
Section \ref{section_six}. 

\section{\.{Z}ukowski's Argument Simplified} \label{section_two}

In effect, \.{Z}ukowski considers the following pair of $\pm 1$-valued
functions of Alice's Bell operator: 
\begin{eqnarray}\nonumber
A_{1} & = & \left| \Psi _{-}\right\rangle \left\langle \Psi
_{-}\right| +\left| \Phi _{-}\right\rangle \left\langle \Phi
_{-}\right| \\ & & -\left| \Psi _{+}\right\rangle \left\langle \Psi
_{+}\right| -\left| \Phi _{+}\right\rangle \left\langle \Phi
_{+}\right| ,  \label{A} \\ \nonumber
A_{2} & = & \left| \Psi _{+}\right\rangle \left\langle \Psi
_{+}\right| +\left| \Phi _{-}\right\rangle \left\langle \Phi
_{-}\right| \\ & & -\left| \Psi _{-}\right\rangle \left\langle \Psi
_{-}\right| -\left| \Phi _{+}\right\rangle \left\langle \Phi
_{+}\right| .  \label{B}
\end{eqnarray}
The operator $A_{1}$ ($A_{2})$ coincides with the observable that represents
the first (second) component of the \textit{vector} observable $%
\overrightarrow{A}$ introduced by \.{Z}ukowski (\cite{z-guy}, Eqn. (12)). It is
easy to see that the information obtained in a joint measurement of $%
A_{1}$ and $A_{2}$ on qubits $1+2$ is equivalent to the
information one would obtain through a single Bell operator measurement. So we
can equally well think of Alice as jointly measuring the observables 
$A_{j}$ ($j=1,2$) in her attempt to teleport the unknown state to Bob.

Now take $D$ to be the maximally entangled state $\left| \Phi
_{+}\right\rangle \left\langle \Phi _{+}\right| $, consider
two alternative unknown states given by 
\begin{eqnarray}
\left| \phi _{1}\right\rangle &=&\frac{1}{\sqrt{2}}(\left| \uparrow
\right\rangle +\left| \downarrow \right\rangle ),  \label{one} \\
\left| \phi _{2}\right\rangle &=&\frac{1}{\sqrt{2}}(\left| \uparrow
\right\rangle +i\left| \downarrow \right\rangle ),\;  \label{two}
\end{eqnarray}
and consider the following two alternative spin components of qubit $3$, 
\begin{eqnarray}
\sigma _{1} &=&e^{-i\pi /4}\left| \uparrow \right\rangle
\left\langle \downarrow \right| +e^{i\pi /4}\left| \downarrow
\right\rangle \left\langle \uparrow \right| ,  \label{three} \\
\sigma _{2} &=&e^{i\pi /4}\left| \uparrow \right\rangle \left\langle
\downarrow \right| +e^{-i\pi /4}\left| \downarrow \right\rangle
\left\langle \uparrow \right| .  \label{four}
\end{eqnarray}
(It is easily checked that the choices in (\ref{one})--(\ref{four})
correspond to the choices \.{Z}ukowski (\cite{z-guy}, p. 2) makes for the parameters
he labels as $\beta ,\phi ,\beta',\phi'$.) \.{Z}ukowski shows
that the following Bell-CHSH-type inequality 
\begin{equation}
\left| 
\begin{array}{c}
\left\langle A_{1}\otimes \sigma _{1}\right\rangle _{\left| \phi
_{1}\right\rangle \left\langle \phi _{1}\right| \otimes
D}+\left\langle A_{1}\otimes \sigma _{2}\right\rangle _{\left|
\phi _{1}\right\rangle \left\langle \phi _{1}\right| \otimes D}
\\ 
+\left\langle A_{2}\otimes \sigma _{1}\right\rangle _{\left| \phi
_{2}\right\rangle \left\langle \phi _{2}\right| \otimes
D}-\left\langle A_{2}\otimes \sigma _{2}\right\rangle _{\left|
\phi _{2}\right\rangle \left\langle \phi _{2}\right| \otimes D}
\end{array}
\right| \leq 2,  \label{Z-expression}
\end{equation}
which we henceforth for convenience call a ``Bell teleportation inequality'', 
is violated by a factor of $\sqrt{2}$. \.{Z}ukowski appears to interpret this 
violation as signifying that the quantum component of a perfect teleportation
process involving the maximally entangled state $D=\left| \Phi
_{+}\right\rangle \left\langle \Phi _{+}\right| $ must be
nonlocal.

This interpretation is not without plausibility. The idea is that to assess
the nonlocality of $D$ \textit{qua} the quantum component of Alice's
teleportation process, we should continue to have Alice measure the Bell
operator but drop the classical channel, her phone call, from consideration.
Although this frustrates any attempt to complete the process by Bob 
performing an appropriate unitary transformation on his qubit, it does
not preclude asking whether what remains of the process can be modelled by a
local hidden variables theory for any conceivable unknown state. One can
imagine, then, that after preparing the quantum channel state, Clare
randomly feeds to Alice one of two unknown states $\left| \phi
_{j}\right\rangle $, and makes a covert agreement with Bob that he
should simply ignore all of Alice's phone calls and randomly measure one of two
spin components $\sigma _{j}$ on his qubit. After many measurement
trials, Clare and Bob then reveal their charade to Alice, who grudgingly
agrees to combine the information from her Bell operator outcomes with their
information to see whether the Bell teleportation inequality (\ref
{Z-expression}) is violated. When they find that it \emph{is} violated, the standard
argument for Bell's theorem (cf. \cite{z-guy}, Eqns. (16),(17)) shows that
no hidden variables model according to which: (i) Alice's and Bob's measurement
outcomes are statistically independent at the level of the hidden 
variables;
(ii) Alice's outcomes are independent of the spin component Bob 
measures; and (iii) \emph{Bob's outcomes are independent of the unknown 
state Alice attempts to teleport}, can
possibly account for the observed correlations.

Right after pointing out that inequality (\ref
{Z-expression}) is violated in the state $D=\left| \Phi
_{+}\right\rangle \left\langle \Phi _{+}\right| $,
\.{Z}ukowski remarks:

\begin{quote}
It is an interesting fact, that needs further investigation, that the Bell
inequality presented here is violated by the same factor $\sqrt{2}$ as the
CHSH inequality for the usual Bell theorem involving a pair of particles in
a maximally entangled state. This may imply that the quantum component of
the teleportation process cannot be described in a local and realistic way,
as long as the initial state of $B$ [$2$] and $C$ [$3$] admits no such model (\cite
{z-guy}, p. 3).
\end{quote}
\.{Z}ukowski's conjecture appears to be that if the channel state violates
{\em any} local hidden variables theory inequality, perhaps even collective
measurement ones, then the correlations involved in the 
quantum component of teleportation will also
violate a local hidden variables theory inequality. If this is indeed the 
conjecture,
it would be difficult to decide, since we would need to take into 
account
all possible local hidden variables theories inequalities for 
all possible measurement protocols on the channel particles alone 
and determine whether a violation of any such inequality suffices to 
prevent the recovery, within a local hidden variables model, of the 
correlations of the channel state that are exploited in the teleportation 
process. 
Consequently, we narrow our focus 
to the standard
Bell-CHSH inequality for the channel state and a generalized form
of (\ref{Z-expression})'s Bell teleportation inequality for qubits 1--3 
where the observables $A_{1}$ and $A_{2}$ are allowed to be \emph{any} 
bivalent functions of Alice's Bell operator.

\section{Violation of a Bell teleportation
inequality implies violation of a Bell-CHSH inequality by the channel 
state}
\label{section_three}

We first generalize \.{Z}ukowski's 
scheme to include a wider class of observables compatible with 
Alice's Bell operator, by 
letting $A_{j}$ ($j=1,2$) be \emph{any} pair of self-adjoint unitary `spin'
operators that are each bivalent functions of her Bell operator.
Thus, 
\begin{eqnarray} \nonumber
A_{j}  & = & a_{j}\left| \Psi _{+}\right\rangle \left\langle \Psi
_{+}\right| +b_{j}\left| \Psi _{-}\right\rangle \left\langle
\Psi _{-}\right| \\ \label{define_A} & & +c_{j}\left| \Phi _{+}\right\rangle \left\langle \Phi
_{+}\right| +d_{j}\left| \Phi _{-}\right\rangle \left\langle
\Phi _{-}\right|,\ j=1,2,
\end{eqnarray}
where $a_{j},b_{j},c_{j},d_{j}=\pm 1$. Now fix, once and for all, the channel
state density operator $D$.  By analogy with 
$\beta(D)$,  define the number $\tau
(D)$ to be the maximum of the Bell teleportation expression 
\begin{equation}
\left| 
\begin{array}{c}
\left\langle A_{1}\otimes \sigma _{1}\right\rangle _{\left| \phi
_{1}\right\rangle \left\langle \phi _{1}\right| \otimes
D}+\left\langle A_{1}\otimes \sigma _{2}\right\rangle _{\left|
\phi _{1}\right\rangle \left\langle \phi _{1}\right| \otimes D}
\\ 
+\left\langle A_{2}\otimes \sigma _{1}\right\rangle _{\left| \phi
_{2}\right\rangle \left\langle \phi _{2}\right| \otimes
D}-\left\langle A_{2}\otimes \sigma _{2}\right\rangle _{\left|
\phi _{2}\right\rangle \left\langle \phi _{2}\right| \otimes D}
\end{array}
\right|  \label{eq:new}
\end{equation}
over all choices of the unknown state vectors $\left| \phi
_{j}\right\rangle $ and all choices of the bivalent Bell operator 
functions $A_{j}$%
, i.e., all $\pm 1$ choices for the numbers 
$a_{j},b_{j},c_{j},d_{j}$.  

Notice that, since there is a unitary 
operator $U$ on the state space of qubit $1$ mapping 
$\left| \phi_{2}\right\rangle$ to $\left| 
\phi_{1}\right\rangle$, (\ref{eq:new}) equals
\begin{equation} \label{gigo}
\left|\left\langle A_{1}\otimes(\sigma _{1}+
\sigma _{2})+
(UA_{2}U^{-1})\otimes(\sigma _{1}-\sigma _{2})
\right\rangle_{\left| \phi_{1}\right\rangle
\left\langle \phi_{1}\right|\otimes D}\right|,
\end{equation}
which is just the absolute value of the expectation value of a Bell-CHSH operator on the bipartite 
system $(1+2)+3$.  Thus, necessarily, $\tau (D)\in 
[2,2\sqrt{2}]$.  Note also that if the unknown states $\left| 
\phi_{1}\right\rangle$ and $\left| \phi_{2}\right\rangle$ are 
compatible (i.e., either the same up to an irrelevant phase or orthogonal), then $U$ can always 
be chosen so that $UA_{2}U^{-1}$ is again a bivalent function of 
the Bell operator.  This is most easily 
seen when $\left| 
\phi_{1}\right\rangle=\left| \uparrow \right\rangle$ and $\left| 
\phi_{2}\right\rangle=\left| \downarrow \right\rangle$, in which case the unitary that simply 
permutes these two $z$-eigenstates maps the four Bell 
operator eigenspaces into each other.  (The general case can be reduced 
to this one by first applying a unitary transformation to the unknown states 
to bring them in line with the $z$-basis.) Thus, when the two 
alternative unknown states 
are compatible, $[A_{1},UA_{2}U^{-1}]=0$.  In that case, it follows 
that expression (\ref{gigo}), and thus (\ref{eq:new}), cannot exceed 
2, and  the corresponding Bell 
teleportation inequality cannot be violated.  This is to be expected, since it is known \cite{hi}
that perfect teleportation of any collection of mutually compatible unknown states can be 
achieved with a separable channel state (all of whose correlations, including 
those involved in the teleportation, 
can be modelled by a local hidden variables theory).  

It is not difficult to show that $\tau(D)=\beta 
(D)=2\sqrt{2}$ if and only if $D$ is a maximally entangled 
pure state (by using the fact that any maximally entangled state is 
related to \.{Z}ukowski's choice $D=\left| \Phi
_{+}\right\rangle \left\langle \Phi _{+}\right| $ by a unitary 
operator on Bob's qubit).
We now concentrate on proving, quite generally,
 that $\beta (D)\geq \tau (D)$.

If $\tau (D)=2$ the claim is trivial, so we suppose $\tau 
(D)>2$.  Thus, there are unknown
states $\left| \phi _{j}\right\rangle $ of qubit 1, bivalent Bell 
operator functions $A_{j}$, and spin components $\sigma
_{j}$ of qubit 3 such that 
\begin{eqnarray} \nonumber
\left| 
\begin{array}{c}
\left\langle A_{1}\otimes \sigma _{1}\right\rangle _{\left| \phi
_{1}\right\rangle \left\langle \phi _{1}\right| \otimes
D}+\left\langle A_{1}\otimes \sigma _{2}\right\rangle _{\left|
\phi _{1}\right\rangle \left\langle \phi _{1}\right| \otimes D}
\\ 
+\left\langle A_{2}\otimes \sigma _{1}\right\rangle _{\left| \phi
_{2}\right\rangle \left\langle \phi _{2}\right| \otimes
D}-\left\langle A_{2}\otimes \sigma _{2}\right\rangle _{\left|
\phi _{2}\right\rangle \left\langle \phi _{2}\right| \otimes D}
\end{array}
\right| \\ =\tau (D)>2.  \label{mon}
\end{eqnarray}
Consider the pair of self-adjoint operators $X_{j}$ on qubit 2's state space
defined via the following matrix elements in the $\sigma _{z}$
eigenbasis: 
\begin{equation} \label{matrix}
\left( 
\begin{array}{ll}
\left\langle \phi _{j}\right| \left\langle \uparrow \right|
A_{j}\left| \phi _{j}\right\rangle \left| \uparrow \right\rangle
& \left\langle \phi _{j}\right| \left\langle \uparrow \right|
A_{j}\left| \phi _{j}\right\rangle \left| \downarrow
\right\rangle \\ 
\left\langle \phi _{j}\right| \left\langle \downarrow \right|
A_{j}\left| \phi _{j}\right\rangle \left| \uparrow \right\rangle
& \left\langle \phi _{j}\right| \left\langle \uparrow \right|
A_{j}\left| \phi _{j}\right\rangle \left| \uparrow \right\rangle
\end{array}
\right) ,\ j=1,2.  \label{friend}
\end{equation}
Since each $A_{j}$ is self-adjoint, the matrices in (\ref{friend}) are
self-adjoint; thus each $X_{j}$ is self-adjoint as well. By linearity, (%
\ref{friend}) implies: 
\begin{equation}
\left\langle v\right| X_{j}\left| w\right\rangle =\left\langle
\phi _{j}\right| \left\langle v\right| A_{j}\left| 
\phi_{j}\right\rangle \left| w\right\rangle,\ j=1,2,
\label{googoogoo}
\end{equation}
for all 
qubit 2 vectors $\left|
v\right\rangle ,\left| w\right\rangle$. In particular, when we set both $\left| v\right\rangle $ and $\left|
w\right\rangle $ equal to any unit eigenvector of $X_{j}$ with
corresponding eigenvalue $\mu _{j}$, it follows that $\left| \mu _{j}\right|
\leq $ $\left\| A_{j}\right\| =1$. Thus each $X_{j}$ is a
self-adjoint \textit{contraction}, i.e., satisfies $\left\| 
X_{j}\right\|
\leq 1$.

Next, multiplying both sides of (\ref{googoogoo}) by $\left\langle
x\right| \sigma _{l}\left| y\right\rangle ,$ with $\left|
x\right\rangle ,\left| y\right\rangle$
arbitrary qubit 3 vectors, we obtain 
\begin{eqnarray} \nonumber
& \left\langle v\right| \left\langle x\right| X_{j}\otimes \sigma
_{l}\left| w\right\rangle \left| y\right\rangle \\ = & \left\langle
\phi _{j}\right| \left\langle v\right| \left\langle x\right|
A_{j}\otimes \sigma _{l}\left| \phi _{j}\right\rangle \left|
w\right\rangle \left| y\right\rangle ,\ j,l=1,2.  \label{yoyo}
\end{eqnarray}
Now, since $D$ is Hermitian, $D=\sum_{k=1}^{4}\lambda _{k}\left|
e_{k}\right\rangle \left\langle e_{k}\right| $, where the $\{
e_{k}\}$ form an orthonormal basis for the 2+3 space. 
Since product vectors span this space, (\ref{yoyo}) implies by linearity 
\begin{eqnarray} \nonumber
\left\langle e_{k}\right| X_{j}\otimes \sigma _{l}\left|
e_{k}\right\rangle =  \left\langle \phi _{j}\right| \left\langle
e_{k}\right| A_{j}\otimes \sigma _{l}\left| \phi
_{j}\right\rangle \left| e_{k}\right\rangle \; \\ k=1,...,4; j,l=1,2.
\label{googoo}
\end{eqnarray}
Multiplying both sides of (\ref{googoo}) by $\lambda _{k}$ and summing over $%
k$ results in: 
\begin{equation}
\left\langle X_{j}\otimes \sigma _{l}\right\rangle
_{D}=\left\langle A_{j}\otimes \sigma _{l}\right\rangle
_{\left| \phi _{j}\right\rangle \left\langle \phi _{j}\right|
\otimes D}\;j,l=1,2.  \label{mouse}
\end{equation}
Combining this with (\ref{mon}) yields 
\begin{equation} 
\left| 
\begin{array}{c}
\left\langle X_{1}\otimes \sigma _{1}\right\rangle
_{D}+\left\langle X_{1}\otimes \sigma _{2}\right\rangle
_{D} \\ 
+\left\langle X_{2}\otimes \sigma _{1}\right\rangle
_{D}-\left\langle X_{2}\otimes \sigma _{2}\right\rangle
_{D}
\end{array}
\right| = \tau (D)>2.  \label{yikes}
\end{equation}
Finally note that, although the $X_{j}$ are in general only self-adjoint
contractions, convexity arguments \cite{3man} show that if a
bipartite state (here, our channel state $D$) violates a 
Bell-CHSH-type
inequality with respect to self-adjoint contractions (as in (\ref{yikes})),
then that state also violates, \textit{by at least the same amount}, a standard
Bell-CHSH inequality with respect to observables {\em all} of which are
self-adjoint unitary spin components. 
This completes the proof that $\beta
(D)\geq \tau (D)$.

An immediate corollary of this result is that there are channel states 
that permit nonclassical teleportation yet violate no Bell 
teleportation inequality.  For example, consider the `Werner state' 
\cite{werner_two}
\begin{equation} \label{w}
D_{W}=\bigl( 1-\frac{1}{\sqrt{2}} \bigr) \frac{1}{4} I \otimes I +
\frac{1}{\sqrt{2}} | \Psi_{-}\rangle \langle \Psi_{-}|.
\end{equation}
Since spin operators, and hence Bell-CHSH operators, are traceless, 
\begin{eqnarray} \nonumber
\beta(D_{W}) & = & \frac{1}{\sqrt{2}}\beta(| \Psi_{-}\rangle \langle 
\Psi_{-}|) \\ & = & \frac{1}{\sqrt{2}}2\sqrt{2}=2\geq\tau(D_{W}),
\end{eqnarray}
which forces $\tau (D_{W})=2$.  On the other hand, the teleportation 
fidelity of $D_{W}$ relative to the standard choice of Bob's unitary 
operators (which, in fact, gives the maximum achievable fidelity) is 
\begin{eqnarray} \nonumber
{\cal F}_{st}(D_{W}) & = & (1-\frac{1}{\sqrt{2}}) \times \frac{1}{2} 
+\frac{1}{\sqrt{2}} \times 1 \\ & = & 
\frac{2}{3}(1+\frac{3\sqrt{2}-2}{8})>\frac{2}{3}.
\end{eqnarray}

\section{A channel state Bell-CHSH violation does not imply a violation
of a Bell teleportation inequality.}
\label{section_four}

We now exhibit a family of channel states $D$ for which the converse 
inequality $\tau
(D)\geq \beta (D)$ fails; indeed, for which $\beta (D)>2$ 
while $\tau (D)=2$.  It follows that if one were to interpret {\.Z}ukowski's conjecture (discussed
 in Section \ref{section_two}) to be that
a channel state Bell-CHSH violation implies a violation
of his original inequality, then this particular 
conjecture is false.   

By the reasoning in the previous section which led to Eqn. 
(\ref{yikes}), we know that the maximum Bell teleportation violation 
given a channel state $D$ is attained by an expression of the 
form $|\langle Z\rangle_{D}|$ where
\begin{equation} \label{Z}
Z= X_{1}\otimes \sigma _{1}+
X_{1}\otimes \sigma_{2}+ X_{2}\otimes \sigma_{1}-
 X_{2}\otimes \sigma_{2}, \label{yikes2}
\end{equation}
and the self-adjoint contractions $X_{j}$ ($j=1,2$) have matrix elements 
given by (\ref{matrix}).  We may calculate these matrix elements 
explicitly by parameterizing the two alternative teleported 
states as
\begin{equation}\label{blob}
\left| \phi_{j}\right\rangle =\sin \theta _{j}\left| \uparrow
\right\rangle +\cos \theta _{j}e^{i\vartheta _{j}}\left| \downarrow
\right\rangle ,\;\theta _{j},\vartheta _{j}\in [0,2\pi ).  
\end{equation}
An
elementary but tedious calculation then shows 
\begin{eqnarray} \label{all_classes}
X_{j} &=&\frac{1}{4}(a_{j}+b_{j}+c_{j}+d_{j})I \nonumber \\
&&+\frac{1}{4}(a_{j}-b_{j}+c_{j}-d_{j})\sin 2\theta _{j}\cos \vartheta
_{j}\sigma _{x} \nonumber \\
&&+\frac{1}{4}(a_{j}+d_{j}-b_{j}-c_{j})\sin 2\theta _{j}\sin \vartheta
_{j}\sigma _{y} \\ \nonumber
&&+\frac{1}{4}(a_{j}+b_{j}-c_{j}-d_{j})\cos 2\theta _{j}\sigma _{z},
\end{eqnarray}
where the numbers $a_{j},b_{j},c_{j},d_{j}=\pm 1$ correspond to 
the different choices that can be made for the bivalent functions 
$A_{j}$ of Alice's Bell 
operator.  The value of $\tau(D)$ is, therefore, given 
by
\begin{equation} \label{max}
\tau(D)=\max_{\theta_{j},\vartheta_{j}\in [0,2\pi 
);a_{j},b_{j},c_{j},d_{j}=\pm 1;\sigma_{j}}
|\langle Z\rangle_{D}|. 
\end{equation}

 Consider, now, the two parameter family of channel density operators given by 
\begin{equation}
D_{\lambda,\alpha}= (1-\lambda)\frac{1}{4} I\otimes I +\lambda
 | \psi_{\alpha} \rangle \langle \psi_{\alpha} |,\ 
 \lambda,\alpha\in [0,1],
\end{equation}
where
\begin{equation} 
| \psi_{\alpha} \rangle = \alpha |+\rangle |+\rangle + 
\sqrt{1-\alpha^{2}}|-\rangle 
|-\rangle_,
\end{equation}
\begin{equation} 
| \pm \rangle = \frac{1 \pm \sqrt{3}}{\sqrt{2(3\pm\sqrt{3})}} 
| \uparrow \rangle  
+ \frac{1+i}{\sqrt{2(3\pm\sqrt{3})}}|\downarrow\rangle.
\end{equation}
For later reference, note that
\begin{equation} \label{num1}
\langle\psi_{\alpha}|I\otimes 
\sigma_{n}|\psi_{\alpha}\rangle=\frac{1}{\sqrt{3}}(2\alpha^{2}-1),\ 
n=x,y,z,
\end{equation}
\begin{eqnarray} \nonumber
\langle\psi_{\alpha}|\sigma_{m}\otimes 
\sigma_{n}|\psi_{\alpha}\rangle=\frac{1}{3}(1+4\alpha\sqrt{1-\alpha^{2}}), 
\\ 
(m,n)=(x,y),(y,x),(z,z), \label{num2}
\end{eqnarray}
\begin{eqnarray}\nonumber
\langle\psi_{\alpha}|\sigma_{m}\otimes 
\sigma_{m}|\psi_{\alpha}\rangle & = &
\langle\psi_{\alpha}|\sigma_{m}\otimes 
\sigma_{z}|\psi_{\alpha}\rangle \\ \label{num3}
& = &
\langle\psi_{\alpha}|\sigma_{z}\otimes 
\sigma_{m}|\psi_{\alpha}\rangle \\ \nonumber & = & \frac{1}{3}
(1-2\alpha\sqrt{1-\alpha^{2}}),\ 
m=x,y.
\end{eqnarray}
Since spin operators are traceless, the corresponding expectation values in the state 
$D_{\lambda,\alpha}$ are obtained by multiplying each of the above values 
by $\lambda$.

Again, because spin operators are traceless, the first maximally mixed component 
of $D_{\lambda,\alpha}$
 does not contribute to any Bell-CHSH operator average; hence 
 $\beta(D_{\lambda,\alpha})>2$ just in case
$\lambda \beta(| \psi_{\alpha} \rangle \langle \psi_{\alpha} |) > 2$.
It can be shown that \cite{popescu_rohrlich} 
\begin{equation}\label{pop}
\beta(| \psi_{\alpha} \rangle \langle 
\psi_{\alpha} |)=2\sqrt{1 + 4 \alpha^{2} (1-\alpha^{2})},  
\end{equation}
thus the condition for 
$\beta(D_{\lambda,\alpha})>2$ is simply 
\begin{equation} \label{eq:bell}
\lambda \sqrt{1 + 4 \alpha^{2} (1-\alpha^{2})} > 1,\ \lambda,\alpha\in [0,1].
\end{equation}
We now determine two additional conditions on $\lambda$ and $\alpha$, 
\emph{consistent} with (\ref{eq:bell}), 
which together are sufficient  for $\tau(D_{\lambda,\alpha}) = 2$.

The set
of 16 possible value assignments to the numbers $a_{j},b_{j},c_{j},d_{j}$ 
can be divided into three relevant classes.  \textbf{Class I}: All 
assignments for which both $A_{1}$ and $A_{2}$
involve two value assignments of both signs (for example, 
the two value assignments $a_{j}=b_{j}=-c_{j}=-d_{j}=-1$,
where $j=1,2$).  \textbf{Class II}: All assignments
 for which exactly one of $A_{1}$ and $A_{2}$
involve two value assignments of both signs and the other
involves three values assignments of one sign (for example,
$a_{1}=b_{1}=-c_{1}=-d_{1}=-1$ and $a_{2}=b_{2}=c_{2}=-d_{2}=-1$). 
\textbf{Class III}: All assignments for which both 
$A_{1}$ and $A_{2}$
involve exactly three values assignments of one sign (for example,
the value assignments 
$a_{j}=b_{j}=c_{j}=-d_{j}=-1$).  Note that
whenever $a_{j} = b_{j} = c_{j} = d_{j}$ for at least one 
value of $j$---i.e., whenever either $A_{1}$ or $A_{2}$ is 
multiple of the identity---no violation is possible, thus such assignments
can be ignored.

\textbf{Class I}: Following \cite{popescu_rohrlich}, we introduce the 
new spin operators 
\begin{equation} \label{new_ops}
{\hat \sigma}_{1}=  \frac{\sigma_{1}
 + \sigma_{2}}{2 \cos \xi},\  
{\hat \sigma}_{2}= \frac{\sigma_{1} - \sigma_{2}}{2 \sin \xi},
\end{equation}
so that (\ref{yikes2}) becomes
\begin{equation}
Z=2 (X_{1} \otimes {\hat \sigma}_{1})\cos \xi  + 
2 (X_{2} \otimes {\hat \sigma}_{2}) \sin \xi. 
\end{equation}
Using the fact that, whenever $M,N\in \mathbb{R}$,
$M \cos \xi + N  \sin \xi \leq \sqrt{M^{2} + N^{2}}$, we have 
\begin{equation} \label{zee}
|\langle Z \rangle_{D_{\lambda,\alpha}}|
\leq 2 \sqrt{\langle X_{1} \otimes {\hat \sigma}_{1} 
\rangle_{D_{\lambda,\alpha}}^{2}
+ \langle X_{2} \otimes {\hat \sigma}_{2} \rangle_{D_{\lambda,\alpha}}^{2}}. 
\end{equation}
 Using similar reasoning,
\begin{equation}
\langle X_{j} \otimes {\hat \sigma}_{j}\rangle_{D_{\lambda,\alpha}} 
\leq \sqrt{ \sum_{n=x,y,z}\langle X_{j}\otimes \sigma_{n} 
\rangle_{D_{\lambda,\alpha}}
^{2}} ,\ j=1,2,
\end{equation}
therefore
\begin{equation} \label{temp}
|\langle Z\rangle_{D_{\lambda,\alpha}}| \leq 2\sqrt{ \sum_{n=x,y,z;j=1,2}\langle X_{j}\otimes \sigma_{n} 
\rangle_{D_{\lambda,\alpha}}
^{2}}. 
\end{equation}
Assuming a \textbf{Class I} value assignment to the numbers 
$a_{j},b_{j},c_{j},d_{j}$, each self-adjoint contraction in 
(\ref{all_classes}) has the form 
$X_{j} = r_{j}\sigma_{m_{j}}$ where $r_{j}\in[-1,1]$ and 
$m_{j}=x,y,z$.  Since $|\langle Z\rangle_{D_{\lambda,\alpha}}|$ 
cannot exceed the classical bound $2$ when $[X_{1},X_{2}]=0$, we need 
only consider the three cases where 
$m_{1}\not=m_{2}$, i.e.,
\begin{eqnarray}\label{in} 
X_{1}=r_{1} \sigma_{x}, & X_{2}=r_{2} \sigma_{y}, \\ 
X_{1}=r_{1} \sigma_{y}, & X_{2}=r_{2} \sigma_{z}, \\ 
X_{1}=r_{1} \sigma_{x}, & X_{2}=r_{2} \sigma_{z}, 
\end{eqnarray} 
and the other three cases obtained by interchanging 1 and 2.  
For the first case, 
substitution into inequality (\ref{temp}) and using the expectation 
values given in (\ref{num2}) and (\ref{num3}), we find
\begin{eqnarray} \nonumber
|\langle Z \rangle _{D_{\lambda,\alpha}} | & \leq  & 
\frac{2 \lambda}{3} \sqrt{(r_{1}^{2} + r_{2}^{2})
( 3+24\alpha^{2} (1-\alpha^{2}))} \\ \label{ineq}
& \leq & \frac{2  \sqrt{2} \lambda}{\sqrt{3}} \sqrt{1+8\alpha^{2} (1-\alpha^{2})}.
\end{eqnarray}
The other two cases in (\ref{in}) yield the same inequality 
(\ref{ineq}), as do the cases obtained by interchanging 1 and 2 
(which leaves the right-hand side of inequality (\ref{temp}) unchanged).   
Thus, a sufficient condition for no \textbf{Class I} value 
assignment to yield a violation of a Bell teleportation inequality is 
\begin{equation} \label{class_one}
\lambda \sqrt{2/3\bigl( 1 + 8\alpha^{2} (1-\alpha^{2})\bigr)} \leq 1.
\end{equation}

\textbf{Class II}: In this case, we can see from (\ref{all_classes}) 
that exactly one of $X_{1}$ and $X_{2}$ 
takes the form $r \sigma_{m}$, where $m=x,y,z$ and $r\in[-1,1]$, while 
the other is of the form
$\pm I/2 + \sigma/2$ where $\sigma$ is a spin-1/2 operator.
Thus, 
\begin{eqnarray} \label{third_case}
Z & = & r \sigma_{m} \otimes 
\bigl( \sigma_{2} + {\bar \sigma}_{2} \bigr) + 
1/2(\sigma \pm I)\otimes \bigl( \sigma_{1}- {\bar \sigma}_{2} \bigr) \\
\nonumber 
& = & \pm I \otimes (\sigma_{1} - {\bar \sigma}_{2})/2
+ r \sigma_{m} \otimes (\sigma_{1} + {\bar \sigma}_{2})/2 + B'/2,
\end{eqnarray}
where ${\bar \sigma}_{2}= \pm \sigma_{2}$ and
$B'$ is a Bell-CHSH-type operator constructed out of self-adjoint 
contractions that are not necessarily unitary (unless $r=\pm 1$).  Transforming the 
spin operators $(\sigma_{1},{\bar \sigma}_{2})\mapsto 
(\hat{\sigma}_{1},\hat{\sigma}_{2})$ via Eqns. (\ref{new_ops}),
 and substituting into 
Eqn.~(\ref{third_case}) yields
\begin{equation} \label{not}
Z= \pm  (I \otimes {\hat \sigma}_{2})  
\sin \xi
+ r (\sigma_{m} \otimes {\hat \sigma}_{1}) 
\cos \xi
+ B'/2.
\end{equation}
Therefore,
\begin{eqnarray} \nonumber
& | \langle Z \rangle_{D_{\lambda,\alpha}} | \\ \nonumber 
\leq & | \pm \langle I \otimes {\hat \sigma}_{2}\rangle_{D_{\lambda,\alpha}} 
\sin \xi \\ \nonumber & +
r \langle \sigma_{m} \otimes {\hat \sigma}_{1} \rangle_{D_{\lambda,\alpha}} 
\cos \xi | 
+ 1/2 | \langle B' \rangle_{D_{\lambda,\alpha}} | \\ \label{partial} 
\leq & \sqrt{ \langle I \otimes {\hat \sigma}_{2} 
\rangle_{D_{\lambda,\alpha}}^{2} +
r^{2} \langle \sigma_{m} \otimes \hat{\sigma_{1}}
\rangle_{D_{\lambda,\alpha}}^{2} } \\ \nonumber & 
+ 1/2 | \langle B' \rangle_{D_{\lambda,\alpha}} |.  
\end{eqnarray}
>From (\ref{pop}) it follows that 
\begin{equation}
| \langle B' \rangle_{D_{\lambda,\alpha}} | \leq 
2\lambda \sqrt{1 + 4 \alpha^{2} (1-\alpha^{2})}
\end{equation}
 and, clearly,
\begin{equation}
r^{2}\langle \sigma_{m} \otimes {\hat \sigma}_{1} \rangle_{D_{\lambda,\alpha}}^{2}
= \lambda^{2} r^{2}\langle \sigma_{m} \otimes {\hat \sigma}_{1} 
\rangle_{| \psi_{\alpha}\rangle\langle \psi_{\alpha}| } 
\leq \lambda^{2}.
\end{equation}
  Moreover, 
\begin{eqnarray}\nonumber
& \langle I \otimes \hat{\sigma}_{2}\rangle^{2}_{D_{\lambda,\alpha}} 
\\ \leq
& \langle I \otimes \sigma_{x} \rangle_{D_{\lambda,\alpha}}^{2} 
+ \langle I \otimes \sigma_{y} 
\rangle_{D_{\lambda,\alpha}}^{2}
+ \langle I \otimes \sigma_{z} \rangle_{D_{\lambda,\alpha}}^{2} 
\\ \nonumber = &\lambda^{2} 
(2 \alpha^{2} - 1)^{2},
\end{eqnarray}
 using Eqn. (\ref{num1}) for the last equality.
Upon substitution of these last three inequalities back into inequality
(\ref{partial}), we 
obtain
\begin{eqnarray}\nonumber
|\langle Z \rangle_{D_{\lambda,\alpha}}| & \leq & 
\lambda \bigl(\sqrt{2(1-2\alpha^{2} (1-\alpha^{2}))} \\ & & + 
\sqrt{1 + 4\alpha^{2} (1-\alpha^{2})} \bigr) .
\end{eqnarray}
Thus, a sufficient condition for no \textbf{Class II} value 
assignment to yield a violation of a Bell teleportation inequality is 
\begin{equation} \label{class_three}
\lambda \bigl(\sqrt{2(1-2\alpha^{2} (1-\alpha^{2}))} + 
\sqrt{1 + 4\alpha^{2} (1-\alpha^{2})} \bigr)\leq 2.
\end{equation}

\textbf{Class III}: In this case, $Z$ takes the form
$I \otimes \sigma + B/2$ where $\sigma$ is a spin-operator and $B$ a 
standard Bell-CHSH operator.  Note that this is just a special case of 
the \textbf{Class II} expression for $Z$ in Eqn. (\ref{not}), taking 
 the $\pm$ sign to be $+$, 
$\hat{\sigma}_{2}=\sigma$, $\xi=\pi/2$,
 and $B' = B$. Thus inequality 
(\ref{class_three}) is also sufficient for no \textbf{Class III} value 
assignment to yield a violation of a Bell teleportation inequality.

Combining our results for each class, we see that $\beta (D_{\lambda,\alpha})>2$ 
while $\tau (D_{\lambda,\alpha})=2$ if and only if $\lambda$ and 
$\alpha$ satisfy conditions 
(\ref{eq:bell}),
(\ref{class_one}) and (\ref{class_three}).  The family of channel density operators 
satisfying these conditions is indeed nonempty.  For example, the 
conditions are 
satisfied by taking $\lambda = \sqrt{3/5}\approx .77$ and  
$\alpha = \sqrt{3}/2\approx .87$, and in that case we have 
$\beta(D_{\sqrt{3/5},\sqrt{3}/2})=2\sqrt{21/20}\approx 2.05$.  Note, also, that choosing $\lambda=1$ 
does \emph{not} satisfy the conditions, nor does choosing $\alpha^{2} 
(1-\alpha^{2})=1/4$.  Thus no channel density operator in the family 
is pure, and the pure entangled component 
$| \psi_{\alpha} \rangle \langle \psi_{\alpha} |$ of 
$D_{\lambda,\alpha}$ is never maximally entangled. Finally, since 
$\beta (D)>2$ implies ${\cal F}_{max}(D)>2/3$ \cite{horodeckis}, each 
density operator in the family supplies another example of a state 
that permits nonclassical teleportation without violating any Bell 
teleportation inequality.

\section{Fidelity values that imply a violation
of a Bell teleportation inequality.}
\label{section_five}

In light of our previous results, it is natural to ask what values of 
the fidelity are sufficient for the quantum correlations 
in the standard teleportation protocol \emph{not} to admit a local hidden variables 
explanation.  We shall show in 
this section that ${\cal F}_{st}(D)>2/3(1+1/2\sqrt{2})\approx .90$  implies 
$\tau (D)>2$.
   
It is known \cite{horodecki} that for any Bell-CHSH operator $B$,
\begin{equation}
\langle B \rangle_{D} = \mathbf{a}\cdot T(D)( \mathbf{b}+\mathbf{b'} )
+ \mathbf{a'}\cdot T(D)( \mathbf{b}-\mathbf{b'} )
\end{equation}
where $\mathbf{a}, \mathbf{a'}, \mathbf{b}$ and $\mathbf{b'}$
are unit vectors in $\mbox{$\mathbb{R}$}^{3}$ defining the four spin operators that occur in
$B$, and the $3 \times 3$ matrix $T(D)$ has components
\begin{equation}
T_{mn}(D)=\mbox{Tr}[D(\sigma_{m}\otimes\sigma_{n})],\ m,n=x,y,z,
\end{equation}
encoding the inter-particle correlations in state $D$.
Setting $\mathbf{a}=\mathbf{x}$, $\mathbf{a'}=\mathbf{y}$,
$\mathbf{b}=(\mathbf{x} + \mathbf{y})/\sqrt{2}$,
and $\mathbf{b'}=(\mathbf{x} - \mathbf{y})/\sqrt{2}$,
we obtain 
\begin{eqnarray} \nonumber
\langle B \rangle_{D} & = & 
\langle \sigma_{x}\otimes\sigma_{b}+\sigma_{x}\otimes\sigma_{b'}+
\sigma_{y}\otimes\sigma_{b}-\sigma_{y}\otimes\sigma_{b'} \rangle_{D} 
\\ \label{T_bell}
& = & \sqrt{2} (T_{xx}(D) + T_{yy}(D)).
\end{eqnarray}
Now choose the operators $A_{j}$ in Eqn. (\ref{define_A})
that correspond to $a_{1}=c_{1}=-b_{1}=-d_{1}=+1$
and $a_{2} = d_{2} = -b_{2} = -c_{2} = +1$, and the unknown 
states $\left| \phi_{j}\right\rangle$ in (\ref{blob}) that correspond 
to $\theta_{1}=\pi/4$, $\vartheta_{1}=0$, 
$\theta_{2}=\pi/4$, and $\vartheta_{2}=\pi/4$.  
Substituting these values into (\ref{all_classes}), we  
obtain $X_{1}=\sigma_{x}$ and $X_{2}=\sigma_{y}$.  Hence the Bell-CHSH 
operator $B$ in (\ref{T_bell}) has the form of a $Z$ operator as given in (\ref{Z}), 
and $|\langle B \rangle_{D}|$ provides a lower bound for 
$\tau(D)$, i.e.,  
\begin{equation} \label{bog}
  \tau(D)\geq \sqrt{2} |T_{xx}(D) + T_{yy}(D)|.
  \end{equation}
  (As check on the correctness of this bound, note from (\ref{num3}), that it 
  falls below 2 when $D=D_{.77,.87}$, as it must.)  

It can be shown quite generally that \cite{horodeckis}
\begin{equation} \label{fidelity}
{\cal F}_{st}(D) =  \frac{1}{8} \sum_{n=1}^{4} (1 + \frac{1}{3} 
\mbox{Tr}[T_{n}^{\dag} T(D) O_{n}])
\end{equation}
where 
\begin{eqnarray}
T_{1} = \mbox{diag}(-1,-1,-1), & 
T_{2} = \mbox{diag}(-1,1,1), \\ T_{3} = \mbox{diag}(1,-1,1), & 
T_{4} = \mbox{diag}(1,1,-1),
\end{eqnarray}
 and the matrices $\{O_{n}\}$ are 
determined by the standard choice for Bob's unitary operations in 
(\ref{standard}) via the requirement that
$U_{n}(\mathbf{b}\cdot\mathbf{\sigma})U_{n}^{-1}=(O_{n}^{\dag}\mathbf{b})\cdot
\mathbf{\sigma}$ for all $\mathbf{b}\in \mbox{$\mathbb{R}$}^{3}$.  
Calculating out these latter matrices explicitly yields
\begin{eqnarray}
O_{1}=\mbox{diag}(1,1,1), & O_{2}=\mbox{diag}(1,-1,-1), \\ 
O_{3}=\mbox{diag}(-1,1,-1), & O_{4}=\mbox{diag}(-1,-1,1).
\end{eqnarray}
(For example, for $n=2$, $U_{2}=\sigma_{x}$, and, using the fact 
that $\sigma_{x}^{2}=I$ and 
orthogonal spin components anti-commute,
\begin{eqnarray}
\sigma_{x}(\mathbf{b}\cdot\mathbf{\sigma})\sigma_{x} 
& = & 
\sigma_{x}(b_{x}\sigma_{x}+b_{y}\sigma_{y}+b_{z}\sigma_{z})\sigma_{x} \\
& = & b_{x}\sigma_{x}-b_{y}\sigma_{y}-b_{z}\sigma_{z} \\ & = &
(O_{2}^{\dag}\mathbf{b})_{x}\sigma_{x}+(O_{2}^{\dag}\mathbf{b})_{y}\sigma_{y}+
(O_{2}^{\dag}\mathbf{b})_{z}\sigma_{z},
\end{eqnarray}
which implies $O_{2}=\mbox{diag}(1,-1,-1)$.)  
Plugging all the matrices $\{T_{n}\}$ and $\{O_{n}\}$ back into 
(\ref{fidelity}) gives
\begin{equation} \label{gog}
  {\cal F}_{st}(D) =1/2 -1/6(T_{xx}(D) + T_{yy}(D) + T_{zz}(D)).
\end{equation}

Finally, combining (\ref{bog}) with Eqn. (\ref{gog}), we have
\begin{equation}
\tau(D)\geq \sqrt{2} |T_{zz}(D) + 6{\cal F}_{st}(D)-3|.
\end{equation}
Assuming ${\cal F}_{st}(D)>2/3(1+1/2\sqrt{2})=2/3+\sqrt{2}/6$, and noting that 
$T_{zz}(D)\geq -1$, it follows that 
\begin{eqnarray}\nonumber
& \sqrt{2}(T_{zz}(D) + 6{\cal F}_{st}(D)-3) \\ > & \sqrt{2}(-1+4+\sqrt{2}-3)=2,
\end{eqnarray}  
 and therefore 
$\tau(D) > 2$, as claimed.  Note that we have made no attempt to find 
the
 minimum value for ${\cal F}_{st}(D)$
that implies $\tau(D) > 2$.  However, for the Werner state in 
(\ref{w}), 
we know that ${\cal F}_{st}(D_{W})= 
2/3(1+(3\sqrt{2}-2)/8)\approx .85$, yet $\tau(D_{W})=2$; so our bound of 
$.90$ cannot be 
decreased below
$.85$.

\section{Discussion}\label{section_six}

We have compared a new class of ``Bell teleportation inequalities''---our 
generalization of {\.Z}ukowski's inequality---to
the well-known class of Bell-CHSH inequalities for the particle pair that 
makes up the quantum channel of a standard teleportation process.
We found that while a Bell teleportation inequality cannot be 
violated unless the channel state already violates a Bell-CHSH inequality, 
the latter is no guarantee that the correlations of the channel state 
actually involved in the 
teleportation process will violate a Bell teleportation inequality.  
This suggests that it is generally \emph{easier} for a local hidden variables 
theory to simulate teleportation than to simulate the results of a 
standard Bell correlation experiment on the channel particles alone.  
Moreover, if one were to interpret {\.Z}ukowski's conjecture (discussed
 in Section \ref{section_two}) to be that
a channel state Bell-CHSH violation implies a violation
of his original inequality, then we have shown this particular 
conjecture to be false.    

Our results also contribute to the
larger debate (discussed in our introduction)
 over the role that nonlocality plays in teleportation. We 
 showed that a Bell teleportation inequality is
always violated when the fidelity of transmission in the standard teleportation
protocol exceeds the classical limit of $2/3$ by a factor of $1+1/2\sqrt{2}$. 
If one accepts that the violation of an
inequality to which any local hidden variables theory is committed implies the 
presence of nonlocality, it follows that standard teleportation with 
a fidelity exceeding $2/3(1+1/2\sqrt{2})\approx .90$ cannot occur without the 
involvement of nonlocality.  
On the other hand, our results suggest the existence of a range 
$(.67,.90)$ of 
\emph{nonclassical} values 
of the fidelity for which a local hidden variables theory of the 
teleportation process may be possible.  
Thus, the ability of an entangled channel state, such 
as $D_{W}$ or $D_{.77,.87}$, to permit 
nonclassical teleportation cannot by itself suffice for concluding 
that the channel state itself, or the teleportation it facilitates, is nonlocal; attention must also be paid to the 
magnitude of the fidelity achievable using the state.

Our tentative conclusion that local hidden variables models of quantum 
teleportation may exist for nonclassical fidelities up to 
$\approx 0.90$ is compatible with Gisin's 
\cite{gisin} demonstration that the \emph{end results} of quantum 
teleportation can be classically simulated up to a fidelity of 
$\approx 0.87$.  However, this comparison should not be given too 
much weight because 
Gisin's simulation has little to do with what actually goes on in the 
standard quantum teleportation protocol.  In his simulation, Alice 
does not measure any Bell operator on qubits 1+2.  Rather, it is 
assumed that she has full knowledge of the `unknown' state's 
expansion coefficients, and her task is simply to classically communicate to 
Bob as much information as she can about these coefficients by using 
only 2 bits.  While it is certainly of interest to study how well 
teleportation can be achieved classically using various protocols 
different from the standard quantum protocol (see also \cite{gisin2}), we have confined 
ourselves in this paper to the possibility of local classical 
explanations of the latter only.  While we have reached a negative 
conclusion 
for fidelities exceeding $\approx .90$, our results strongly 
suggest a positive answer for fidelities in the 
range 
$(.67,.90)$. 

\
\vskip10pt
\centerline{\bf Acknowledgements}
\vskip5pt
DTP acknowledges the support
of a Graduate School Research Travelling Award from the University 
of Queensland and thanks the 
Department of Philosophy at the University of Pittsburgh for its 
hospitality.  We would also like to thank Peter Drummond, Gerard
Milburn, and William Munro for helpful discussion.

\end{multicols}


\begin{thebibliography}{99}

\bibitem{popescu_rohrlich} S. Popescu, D. Rohrlich, Phys. Lett. A 166 (1992) 293.

\bibitem{6man}  C. H. Bennett, G. Brassard, C. Cr\'{e}peau, R. Jozsa, A.
Peres, and W. K. Wooters, Phys. Rev. Lett. 70 (1993) 1895.

\bibitem{bmr} S. L. Braunstein, A. Mann, M. Revzen, Phys. Rev. Lett. 68
(1992) 3259.

\bibitem{werner}  R. F. Werner, \texttt{quant-ph/0003070}.

\bibitem{popescu}  S. Popescu, Phys. Rev. Lett. 72 (1994) 797.

\bibitem{gisin}  N. Gisin, Phys. Lett. A 210 (1996) 157.

\bibitem{horodeckis}  R. Horodecki, M. Horodecki, and P. Horodecki, 
\texttt{\ quant-ph/9606027.}

\bibitem{massar}  S. Massar and S. Popescu, Phys. Rev. Lett. 74 (1995) 1259.

\bibitem{vaidman}  L. Vaidman, \texttt{quant-ph/9810089.}

\bibitem{bub}  J. Bub, Stud. Hist. Phil. Mod. Phys. 31 (2000) 74.

\bibitem{bohm}  O. Maroney and B. J. Hiley, Found. Phys. 29 (1999) 1403.

\bibitem{hardy}  L. Hardy, \texttt{quant-ph/9906123.}

\bibitem{deutsch}  D. Deutsch and P. Hayden, \texttt{quant-ph/9906007.}

\bibitem{ari}  A. Duwell, ``Explaining Information Transfer in Quantum
Teleportation'', Phil. Sci. (2001) forthcoming.

\bibitem{z-guy}  M. \.{Z}ukowski, Phys. Rev. A 62 (2000) 032101.

\bibitem{hi} S. Ghosh, G. Kar, A. Roy, and U. Sen, 
\texttt{quant-ph/0010012}. 


\bibitem{3man}  
R. Clifton, H. Halvorson, and A. Kent, 
Phys. Rev. A 61 (2000) 042101.

\bibitem{werner_two} R. F Werner, Phys. Rev. A, 40 (1989) 4277.


\bibitem{horodecki} R. Horodecki, P. Horodecki, and M. Horodecki, 
Phys. Lett. A 200 (1995) 340.

\bibitem{gisin2} N. Cerf, N. Gisin, and S. Massar, Phys. Rev. Lett., 
84 (2000) 2521.

\end{thebibliography}
\end{document}